\documentclass[twocolumn,letter]{jpsj3}

\usepackage{color}
\usepackage{graphicx}% Include figure files
\usepackage{dcolumn}% Align table columns on decimal point
\usepackage{bm}% bold math

\begin{document}
\newcommand{\red}[1]{\textcolor{red}{#1}}

\title{Variation in Superconducting Symmetry against Pressure on Noncentrosymmetric Superconductor Cd$_2$Re$_2$O$_7$ Revealed by $^{185/187}$Re Nuclear Quadrupole Resonance}

\author{Shunsaku~Kitagawa$^{1,}$\thanks{E-mail address: kitagawa.shunsaku.8u@kyoto-u.ac.jp}, 
Kenji~Ishida$^{1}$,
Tatsuo~C.~Kobayashi$^{2}$,
Yasuhito~Matsubayashi$^{3,}$\thanks{Present address: Advanced Coating Technology Research Center, National Institute of Advanced Industrial Science and Technology, Tsukuba Central 5, 1-1-1 Higashi, Tsukuba, Ibaraki, Japan},
Daigorou~Hirai$^{3}$,
Zenji~Hiroi$^{3}$}
\inst{$^1$Department of Physics, Graduate School of Science, Kyoto University, Kyoto 606-8502, Japan \\
$^2$Graduate School of Natural Science and Technology, Okayama University, Okayama 700-8530, Japan \\
$^3$Institute for Solid State Physics, University of Tokyo, Kashiwa, Chiba 277-8581, Japan
}

\date{\today}

\abst{
We performed $^{185/187}$Re nuclear quadrupole resonance (NQR) measurements under pressure to investigate the superconducting properties of noncentrosymmetric superconductor Cd$_{2}$Re$_{2}$O$_{7}$ under various crystal structures.
The pressure dependence of superconducting transition temperature $T_{\rm c}$ determined through ac susceptibility measurements is consistent with the results of previous resistivity measurements [T. C. Kobayashi $et al$., J. Phys. Soc. Jpn. 80, 023715 (2011).]. 
Below 2.2~GPa, in the nuclear spin-lattice relaxation rate $1/T_{1}$, a clear coherence peak was observed just below $T_{\rm c}$, indicating conventional $s$-wave superconductivity.
In contrast, the coherence peak disappears at 3.1~GPa, suggesting a change in superconducting symmetry to the $p$-wave dominant state against pressure.
%NQR spectra broadened with increasing pressure probably because of the inhomogeneous crystal structure and were not observed above 3.9~GPa.
}

%\abovecaptionskip=-5pt
%\belowcaptionskip=-10pt

\maketitle

%\section{Introduction} %% No sections necessary for express letters, letters and short notes
Superconductivity without inversion symmetry in the crystal structure has attracted considerable attention as the absence of inversion symmetry can mix spin-singlet and triplet superconductivity, and it generates an unconventional superconducting(SC) state owing to a strong spin-orbit coupling\cite{M.Sigrist_JMMM_2007}.
In spin-singlet and triplet mixed state, the large upper critical field $H_{\rm c2}$ owing to the absence of Pauli-limiting field, and unusual vortex states such as a helical vortex state are expected\cite{R.P.Kaur_PRL_2005}.
%, and thus, it is important for not only foundamental physics but also application to clarify the SC properties and mechanism.
Although such unconventional SC state has been discussed in CePt$_{3}$Si\cite{E.Bauer_PRL_2004}, Ce$M$Si$_{3}$ ($M$ = Rh and Ir)\cite{N.Kimura_PRL_2005,R.Settai_JPSJ_2011}, UIr\cite{T.Akazawa_JPCM_2004}, and Li$_{2}$Pt$_{3}$B\cite{H.Q.Yuan_PRL_2006}, the relationship between the noncentrosymmetric crystal structure and the SC state is still unclear.

\begin{figure}[!tb]
\vspace*{10pt}
\begin{center}
\includegraphics[width=8.5cm,clip]{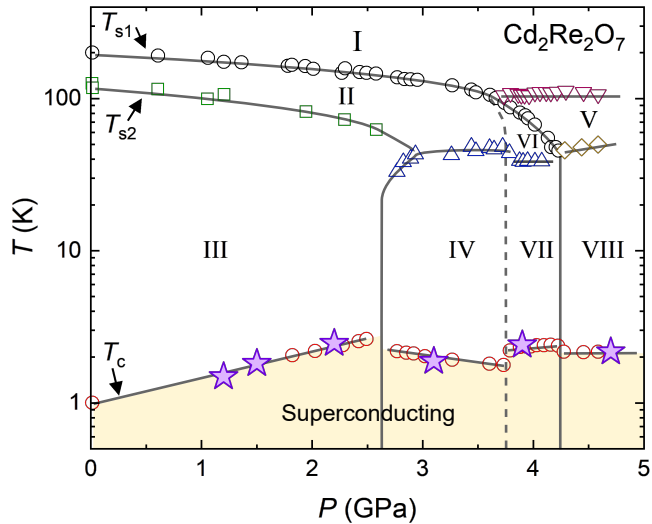}
\end{center}
\caption{(Color online) $P$--$T$ phase diagram of Cd$_{2}$Re$_{2}$O$_{7}$\cite{T.C.Kobayashi_JPSJ_2011,Z.Hiroi_JPSJ_2018}.
Transition temperatures were determined through thermopower and resistivity measurements.
The star symbols represent superconducting (SC) transition determined by our ac susceptibility using a nuclear quadrupole resonance (NQR) coil.
All the lines are visual guides to show possible phase boundaries.
Roman numerals indicate each phase.
}
\label{Fig.1}
\end{figure}

To investigate such a relationship, a pyrochlore oxide Cd$_{2}$Re$_{2}$O$_{7}$ is a suitable material.
Cd$_{2}$Re$_{2}$O$_{7}$ has a cubic structure with space group $Fd\bar{3}m$ (No. 227 $O^7_h$) at room temperature.
Upon cooling, Cd$_{2}$Re$_{2}$O$_{7}$ shows two structural phase transitions at $T_{\rm s1}~\sim$ 200~K (cubic $Fd\bar{3}m$ $\rightarrow$ tetragonal $I\bar{4}m2$) and $T_{\rm s2}~\sim$ 120~K (tetragonal $I\bar{4}m2$ $\rightarrow$ tetragonal $I4_122$), and shows superconductivity below SC transition temperature $T_{\rm c}~\sim$ 1.0~K at ambient pressure\cite{M.Hanawa_PRL_2001,H.Sakai_JPCM_2001,R.Jin_PRB_2001,Z.Hiroi_JPSJ_2018}.
As the low-temperature tetragonal phase lacks inversion symmetry in the crystal structure and a spin-orbit coupling should be strong in such a 5$d$ band metal, unconventional superconductivity was expected in Cd$_{2}$Re$_{2}$O$_{7}$.
However, the SC symmetry of Cd$_{2}$Re$_{2}$O$_{7}$ at ambient pressure is considered to be a conventional $s$-wave based on the experimental results of heat capacity\cite{R.Jin_PRB_2001,Z.Hiroi_JPCS_2002}, $^{187}$Re nuclear quadrupole resonance (NQR)\cite{O.Vyaselev_PRL_2002}, $^{111}$Cd nuclear magnetic resonance (NMR)\cite{H.Sakai_JPSJ_2004}, $\mu$SR\cite{M.D.Lumsden_PRL_2002,R.Kadono_JPSJ_2002} and point-contact spectroscopy\cite{F.S.Razavi_CJP_2015}.
In contrast, the SC properties drastically varied when pressure was applied.
At 4.0 and 4.5~GPa, $T_{\rm c}$ increases to 2.5 K and $H_{\rm c2}$ increases by $\sim$ 30 times than that at ambient pressure, which exceeds the Pauli-limiting field\cite{T.C.Kobayashi_JPSJ_2011}.
Such a large $H_{\rm c2}$ appears to be a hallmark of spin-triplet dominant superconductivity.
In addition, the presence of various structural phases was found experimentally in Cd$_{2}$Re$_{2}$O$_{7}$ under pressure as shown in Fig.~\ref{Fig.1}.
High-pressure X-ray diffraction measurements revealed that phases I, V, and VIII possess inversion symmetry in the crystal structure, and thus, the critical pressure of inversion symmetry breaking order is located at 4.2~GPa\cite{J.Yamaura_PRB_2017}.
These results indicate a close relationship between the inversion symmetry breaking and the enhancement of $H_{\rm c2}$.

%%%%%%%%%%%%%%%%%%%%%%%%%%% Figure 1 %%%%%%%%%%%%%%%%%%%%%%%%%%%%%%%%%%%%%
%%%%%%%%%%%%%%%%%%%%%%%%%%%%%%%%%%%%%%%%%%%%%%%%%%%%%%%%%%%%%%%%%%%%%%%%%%%

Recently, theoretical validation was achieved for such a close relationship\cite{L.Fu_PRL_2015,V.Kozii_PRL_2015,Y.Wang_PRB_2016}. 
According to these theories, a metal with inversion symmetry and strong spin-orbit coupling has a certain Fermi liquid instability that causes inversion symmetry breaking order and the pairing interaction of spin-triplet $p$-wave superconductivity becomes dominant near the quantum critical point of the inversion symmetry breaking order.
They suggested that Cd$_{2}$Re$_{2}$O$_{7}$ is a suitable target to study the proposed intriguing physics because Cd$_{2}$Re$_{2}$O$_{7}$ is a unique material that loses inversion symmetry via the structural phase transition.
Therefore, owing to lack of existing research, investigating the SC properties of Cd$_{2}$Re$_{2}$O$_{7}$ under pressure is important.

Herein, we have performed $^{185/187}$Re NQR measurements to investigate the SC symmetry of Cd$_{2}$Re$_{2}$O$_{7}$ under pressure.
We also measured the pressure dependence of $T_{\rm c}$ determined via ac susceptibility measurements using an NQR coil, which reproduces the results of previous resistivity measurements\cite{T.C.Kobayashi_JPSJ_2011}.
NQR spectra broaden with increasing pressure probably because of the inhomogeneous crystal structure, and they were not observed above 3.9~GPa.
In phase III, a clear coherence peak was observed just below $T_{\rm c}$, indicating conventional $s$-wave superconductivity similar to the result at ambient pressure.
In contrast, a coherence peak disappears at 3.1~GPa (phase IV), indicating that the SC properties are changed owing to a parity mixing.

%\section{Experimental}
High-quality single crystals of Cd$_{2}$Re$_{2}$O$_{7}$ were grown via the chemical transport method, the details of which will be reported elsewhere\cite{Y.Matsubayashi_sample_20xx}.
Single-crystal samples were coarsely ground into a powder to obtain NQR signals with stronger intensity and avoid Joule-heating by radio frequency (RF) pulse for NQR measurements.
$T_{\rm c}$ was determined through ac susceptibility measurements using an NQR coil.
The pressure was generated in an opposed-anvil high-pressure cell designed by K. Kitagawa $et~al.$\cite{K.Kitagawa_JPSJ_2010} with Daphne 7575 as the pressure medium.
The applied pressure $P$ was determined from $T_{\rm c}$ of the lead manometer using the relation of $P$ (GPa) $= [T_{\rm c}(0) - T_{\rm c}(P)]$ (K)/0.364(K/GPa)\cite{A.Eiling_JPFMP_1981,B.Bireckoven_JPESI_1988}.
A conventional spin-echo technique was used for NQR measurements. 
$^{185/187}$Re NQR spectra were obtained as a function of frequency under zero field.
The properties of two isotopes of Re are summarized in Table~\ref{Tab.1}.
A nuclear spin-lattice relaxation rate $1/T_1$ was determined by fitting the time variation of the spin-echo intensity after the saturation of the nuclear magnetization to a theoretical function for $I$ = 5/2.

%%%%%%%%%%%% Table 1 %%%%%%%%%%%%%%%%%%%%%%%%
%\begin{fulltable}[thp]
\begin{table}[bp]
\caption[]{Data of Re isotopes; the nuclear gyromagnetic ratio: $\gamma_n$; the nuclear quadrupolar moment: $Q$; natural abundance: N.A.; and the nuclear spin: $I$.} 
\label{table:Sb_parameter}
%\begin{fulltabular}{rcccc}\hline
\vspace{1cm}
\begin{tabular}{rccccc}
\hline
                   & $\gamma_n/2\pi$(MHz/T) & $Q$(10$^{-24}$cm$^2$ ) & N.A.(\%)  &  $I$\\ \hline
$^{185}$Re & 9.5901 & +2.18 & 37.5 & 5/2 \\
$^{187}$Re & 9.6868 & +2.07 & 62.5 & 5/2  \\
\hline
%\end{fulltabular}
\end{tabular}
\label{Tab.1}
%\end{fulltable}
\end{table}
%%%%%%%%%%%%%%%%%%%%%%%%%%%%%%%%%%%%%%%%%%%%%%%%%

\begin{figure}[!tb]
\vspace*{10pt}
\begin{center}
\includegraphics[width=7.5cm,clip]{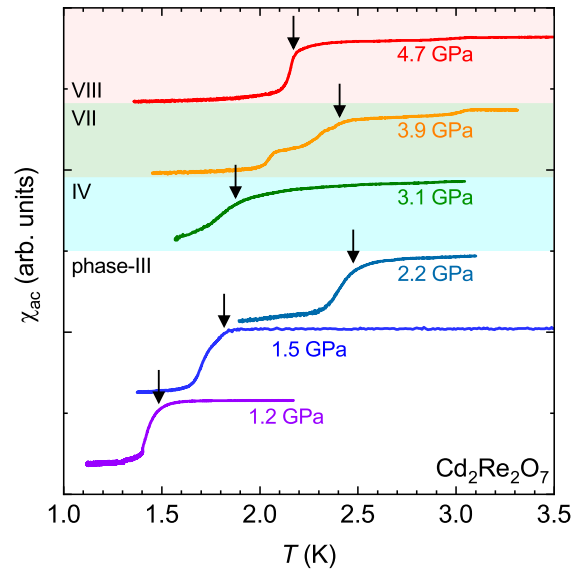}
\end{center}
\caption{(Color online) Temperature dependence of ac susceptibility at zero field under different pressure.
To avoid overlapping data, the offset value is added.
The arrows indicate $T_{\rm c}$ determined by the onset of the diamagnetic signal.
}
\label{Fig.2}
\end{figure}

%\section{Results and Discussion}
First, we checked the pressure dependence of $T_{\rm c}$ in our samples.
Figure~\ref{Fig.2} shows the temperature dependence of ac susceptibility at zero fields under different pressure.
$T_{\rm c}$ was determined by the onset of the diamagnetic signal and is indicated by arrows.
$T_{\rm c}$ increases up to 2.2~GPa and suddenly decreases upon entry into phase IV.
Our results are quite consistent with previous resistivity measurements as shown in Fig.~\ref{Fig.1}\cite{T.C.Kobayashi_JPSJ_2011}.
At 3.1~GPa, SC transition was broader than that at low pressures and double transition was observed at 3.9~GPa.
As sharp SC transition was observed at 4.7~GPa, this broad or double SC transition does not have an extrinsic origin such as pressure inhomogeneity; rather, it originates from an intrinsic effect.
The double SC transition seems to be the evidence that phase VII is the mixtures of phase IV and VIII as discussed in Ref.\citen{Z.Hiroi_JPSJ_2018}.

%%%%%%%%%%%%%%%%%%%%%%%%%%% Figure 2 %%%%%%%%%%%%%%%%%%%%%%%%%%%%%%%%%%%%%
%%%%%%%%%%%%%%%%%%%%%%%%%%%%%%%%%%%%%%%%%%%%%%%%%%%%%%%%%%%%%%%%%%%%%%%%%%%
%%%%%%%%%%%%%%%%%%%%%%%%%%% Figure 3 %%%%%%%%%%%%%%%%%%%%%%%%%%%%%%%%%%%%%
\begin{figure}[!tb]
\vspace*{10pt}
\begin{center}
\includegraphics[width=8.5cm,clip]{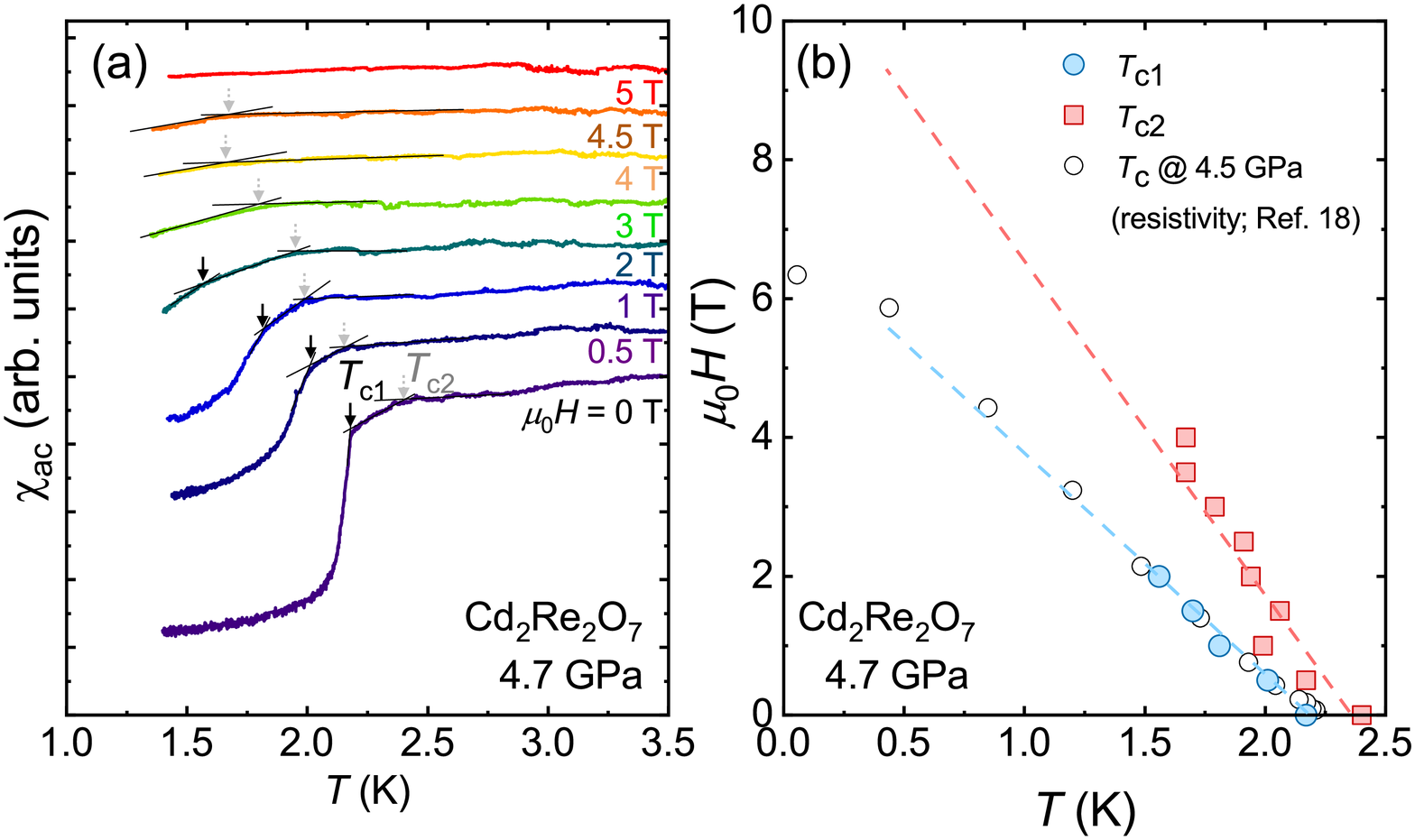}
\end{center}
\caption{(Color online) (a) Temperature dependence of ac susceptibility at 4.7~GPa in different magnetic fields.
$T_{\rm c}$s were determined by the intersection of two extrapolated lines.
The black solid and grey dotted arrows represent $T_{\rm c1}$ and $T_{\rm c2}$,respectively.
(b) $H$--$T$ phase diagram of Cd$_{2}$Re$_{2}$O$_{7}$ at 4.7~GPa.
In addition, $T_{\rm c}$ determined via resistivity measurements at 4.5~GPa is plotted\cite{T.C.Kobayashi_JPSJ_2011}.
Dashed lines are intended for visual guidance.
}
\label{Fig.3}
\end{figure}
%%%%%%%%%%%%%%%%%%%%%%%%%%%%%%%%%%%%%%%%%%%%%%%%%%%%%%%%%%%%%%%%%%%%%%%%%%%
In addition to the peculiar pressure dependence of $T_{\rm c}$ at zero fields, $H_{\rm c2}$ enhancement at 4.7~GPa was reproduced through ac susceptibility measurements.
Figure~\ref{Fig.3} (a) shows the temperature dependence of ac susceptibility at 4.7~GPa in different magnetic fields.
$T_{\rm c}$s determined by the kink of ac susceptibility are plotted against the magnetic field as shown in Fig.~\ref{Fig.3} (b).
Because ac susceptibility shows two kinks, we defined $T_{\rm c1}$ and $T_{\rm c2}$ as a kink at low and high temperatures, respectively.
$T_{\rm c1}$ exhibits almost the same magnetic field dependence as that determined through resistivity measurements at 4.5~GPa\cite{T.C.Kobayashi_JPSJ_2011}.
Notably, $\mu_0H_{\rm c2}~\sim~$6~T exceeds the ordinary Pauli-limiting field $\mu_0H_{\rm P}~\sim~1.84T_{\rm c}~\sim~$4~T, indicating an unconventional SC pairing state.
The high $T_{\rm c}$ ($H_{\rm c2}$) might originate from the surface region because of the small shielding fraction and broad SC transition.
%or the contamination of the high pressure region 

\begin{figure}[!tb]
\vspace*{10pt}
\begin{center}
\includegraphics[width=8.5cm,clip]{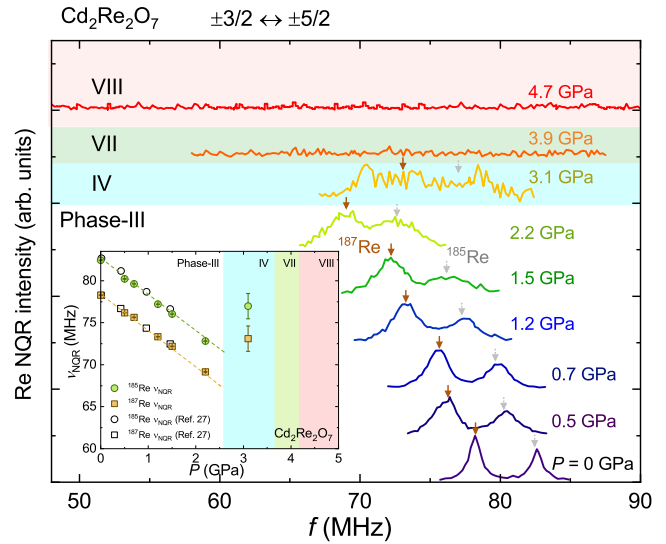}
\end{center}
\caption{(Color online) $^{185/187}$Re NQR spectra of $\pm3/2 \leftrightarrow \pm5/2$ transition at several pressure values in Cd$_{2}$Re$_{2}$O$_{7}$.
The measurement temperature was just above $T_{\rm c}$.
The arrows represent $^{187}$Re and $^{185}$Re NQR frequency $\nu_{\rm NQR}$.
(inset)Pressure dependence of $^{185/187}$Re-$\nu_{\rm NQR}$.
Data from a previous report are also plotted\cite{K.Fujiwara_JPCS_2011}.
Dashed lines are intended for visual guidance.
}
\label{Fig.4}
\end{figure}

Next, we show $^{185/187}$Re NQR results.
Figure~\ref{Fig.4} shows $^{185/187}$Re NQR spectra of $\pm3/2 \leftrightarrow \pm5/2$ transition at several pressure values in Cd$_{2}$Re$_{2}$O$_{7}$.
The NQR spectra were measured just above $T_{\rm c}$.
At ambient pressure, two sharp peaks were observed at 78.2 and 82.6~MHz, which are consistent with those reported in previous works\cite{O.Vyaselev_PRL_2002,K.Fujiwara_JPCS_2011}.
The NQR peak frequency $\nu_{\rm NQR}$ decreases and the linewidth broadens with increasing pressure in Phase III.
The decrease in $\nu_{\rm NQR}$ indicates the suppression of a structural order parameter, as $\nu_{\rm NQR}$ usually increases by applying pressure.
These results are also consistent with those of previous reports\cite{K.Fujiwara_JPCS_2011}.
The pressure dependence of $^{185/187}\nu_{\rm NQR}$ is similar to that in the previous report as shown in the inset of Fig.~\ref{Fig.4}.
Although $^{185/187}$Re NQR spectra disappear at around 2~GPa in the previous report, they were present herein up to 3.1~GPa, indicating better quality of samples and homogeneous pressure in this work.
At 3.1~GPa, $^{185/187}\nu_{\rm NQR}$ suddenly increases and the linewidth significantly broadens owing to entry into phase IV.
With further increasing pressure, the $^{185/187}$Re NQR spectra finally disappears.
As $^{185/187}$Re-NMR spectra were not observed at 4.7~GPa, it is not because of the decrease in $\nu_{\rm NQR}$ but the significant broadening of NQR spectra.
As mentioned above, SC transition is still sharp at 4.7~GPa.
Therefore, it is possible that the broadening of NQR spectra is not due to pressure inhomogeneity accompanied by the solidification of the pressure medium, but rather to intrinsic changes in the distribution of the electric field gradient at the Re nuclear sites.
The structural phase transitions on Cd$_{2}$Re$_{2}$O$_{7}$ are characterized by a large change in $\nu_{\rm NQR}$ with a small change in structural parameters\cite{K.Arai_JPCM_2002}, and thus, a small distribution of the structural parameters may cause the NQR signal to disappear.

%%%%%%%%%%%%%%%%%%%%%%%%%%% Figure 4 %%%%%%%%%%%%%%%%%%%%%%%%%%%%%%%%%%%%%
%%%%%%%%%%%%%%%%%%%%%%%%%%%%%%%%%%%%%%%%%%%%%%%%%%%%%%%%%%%%%%%%%%%%%%%%%%%

\begin{figure}[!tb]
\vspace*{10pt}
\begin{center}
\includegraphics[width=8.5cm,clip]{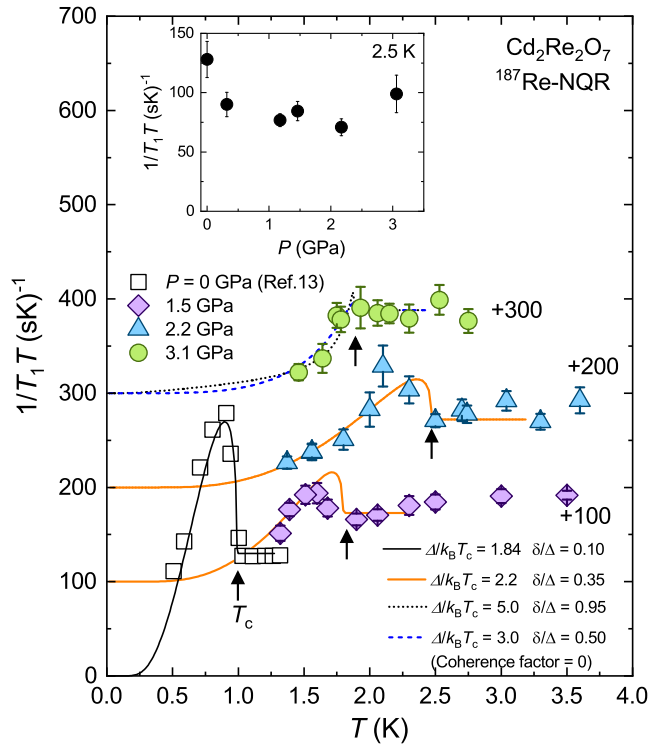}
\end{center}
\caption{(Color online) Temperature dependence of $1/T_1T$ at the $^{187}$Re site up to 3.1~GPa.
Data at ambient pressure are also plotted\cite{O.Vyaselev_PRL_2002}.
To avoid overlapping data, the offset value is added.
The arrows indicate $T_{\rm c}$ determined via ac susceptibility measurements.
Calculations of $1/T_1T$ using the single full-gap model with three different parameters are shown with curves.
The parameters are displayed in the figure.
(inset) Pressure dependence of $1/T_1T$ at 2.5~K.
}
\label{Fig.5}
\end{figure}

As NQR spectra disappear above 3.9~GPa, we measured the temperature dependence of $1/T_1T$ at the $^{187}$Re site up to 3.1~GPa and down to 1.4~K as shown in Fig.~\ref{Fig.5}.
Data at ambient pressure are also plotted for reference\cite{O.Vyaselev_PRL_2002}.
At ambient pressure, $1/T_1T$ was constant at the normal state and a clear coherence peak was observed just below $T_{\rm c}$, strongly evidencing the conventional $s$-wave SC state.
This feature does not vary under pressure up to 2.2~GPa; however, the coherence peak becomes smaller.
The value of $1/T_1T$ in the normal state under pressure (1.5~GPa) is $\sim$ 64~\% of that at ambient pressure as shown in the inset of Fig.~\ref{Fig.5}.
Here, in the normal state, $1/T_1T$ is usuallly proportional to the square of the electronic density of states (DOS), suggesting an decrease in DOS to $\sim$ 80~\%.
The decrease in DOS against pressure is inconsistent with the mass enhancement against pressure deduced from the $A$ coefficient of resistivity and the initial slope of $H_{\rm 2}$\cite{T.C.Kobayashi_JPSJ_2011}.
As $^{185/187}$Re-$1/T_1T$ reflects not only the contribution from DOS but also orbital susceptibility\cite{K.Arai_JPCM_2002}, it is possible that the orbital contribution decreases by applying pressure, owing to the suppression of a structural order parameter.
Note that we did not observe any significant enhancement in $1/T_1T$ upon cooling, which was reported in a previous report\cite{K.Fujiwara_JPCS_2011}.
Because $1/T_1T$ varies across $T_{\rm c}$ in our measurements, the results are deemed accurate, and indicate that the higher-$T_{\rm c}$ superconductivity near the phase boundary is the bulk effect.
In a previous report\cite{K.Fujiwara_JPCS_2011}, they measured the temperature dependence of $1/T_1T$ only above $T_{\rm c}$.
The difference in the results between our and the previous experiments may be caused by the Joule-heating effect.
The Joule-heating effect becomes more significant in pressure cells, and thus, small-energy RF pulses are necessary to obtain accurate results.

%%%%%%%%%%%%%%%%%%%%%%%%%%% Figure 5 %%%%%%%%%%%%%%%%%%%%%%%%%%%%%%%%%%%%%
%%%%%%%%%%%%%%%%%%%%%%%%%%%%%%%%%%%%%%%%%%%%%%%%%%%%%%%%%%%%%%%%%%%%%%%%%%%

In terms of the SC state, the coherence peak was observed up to 2.2~GPa.
In contrast, $1/T_1T$ did not show a coherence peak just below $T_{\rm c}$ at 3.1~GPa.
Several reasons are responsible for suppressing the coherence peak: the existence of gap nodes, the sign change of gap, the spatial distribution of gap size, and large SC gap.
To clarify the origin of the suppression of coherence peak, we analyzed the experimental data.
For simplicity, we assume a single full-gap SC order parameter as observed in the case of ambient pressure.
In this model, the temperature dependence of $1/T_1T$ is expressed as
\begin{align}
\frac{1}{T_1T} \propto &\int^{\infty}_{0} (N_{s}(E)^{2} + M_{s}^{2}) f(E)[1-f(E)]{\rm d}E
\end{align}
with
\begin{align}
N_s(E)/N_0 &= \int^{E+\delta}_{E-\delta} \frac{E'}{\sqrt{E'^2-\Delta^2}}{\rm d}E' \text{, and} \\
M_s(E)/N_0 &= \int^{E+\delta}_{E-\delta} \frac{\Delta}{\sqrt{E'^2-\Delta^2}}{\rm d}E'.
\end{align}
Here, $N_{s}(E), M_{s}(E)$, $N_0$ $\Delta, \delta$, and $f(E)$ are the quasi-particle DOS, the anomalous DOS originating from the coherence effect of Cooper pairs, the DOS in the normal state, the magnitude of SC gap, smearing factor, and the Fermi distribution function, respectively.
The experimental results at 1.5 and 2.2 GPa can be reproduced using $\Delta/k_{\rm B}T_{\rm c}$ = 2.2 and $\delta/\Delta$ = 0.35, as shown in Fig.~\ref{Fig.5}.
Here, $k_{\rm B}$ is the Boltzmann's constant.
The selected parameters are not so different from those at ambient pressure ($\Delta/k_{\rm B}T_{\rm c}$ = 1.84 and $\delta/\Delta$ = 0.10) and can be understood by the variation against pressure.
However, reproducing the experimental results at 3.1~GPa with realistic parameters is difficult.
An unrealistically huge SC gap and gap distribution ($\Delta/k_{\rm B}T_{\rm c}$ = 5.0 and $\delta/\Delta$ = 0.95) are needed to reproduce the temperature dependence of $1/T_1T$ at 3.1~GPa.
Alternatively, if we assume the absence of the coherence factor [$M_{s}(E) = 0$], the experimental results at 3.1~GPa can be reproduced by the proposed model with $\Delta/k_{\rm B}T_{\rm c}$ = 3.0 and $\delta/\Delta$ = 0.50, as shown in Fig.~\ref{Fig.5}.
The larger SC gap is in good agreement with the large $H_{\rm c2}$, and the large smearing factor is consistent with the broad SC transition, as shown in Fig.~\ref{Fig.2}.
The disappearance of the anomalous DOS indicates the sign change of the SC gap, which suggests the pressure-induced $p$-wave dominant SC state.
Therefore, we suggest that the SC gap symmetry changes in phase-IV and it may be related to the enhanced interaction of the spin-triplet channel; however, currently, the experiment probing the pairing symmetry has not been performed yet, determining the SC gap symmetry is difficult.
This change in SC symmetry is consistent with the theoretical phase diagram proposed in Ref.\citen{V.Kozii_PRL_2015}.
Further investigation is required to understand the unconventional SC state under pressure.

%\section{Conclusion}
In conclusion, we performed ac susceptibility and $^{185/187}$Re NQR measurements to investigate the SC symmetry of Cd$_{2}$Re$_{2}$O$_{7}$ under pressure.
The pressure dependence of $T_{\rm c}$ determined via ac susceptibility measurements is consistent with the results of previous resistivity measurements\cite{T.C.Kobayashi_JPSJ_2011}.
Our NQR results suggest that the SC gap symmetry changes from conventional $s$-wave in phase III to $p$-wave dominant state in phase IV with the application of pressure.
This is first evidence suggesting the realization of unconventional SC state in Cd$_{2}$Re$_{2}$O$_{7}$ under pressure from the microscopic point of view.
Our findings promote the further investigation of the properties of spin-triplet superconductivity near the quantum critical point of the inversion symmetry breaking order.

%$^{185/187}$Re NQR spectra broaden with increasing pressure probably because of the intrinsic disorder effect and were not observed above 3.9~GPa.
%According to $1/T_1T$ measurements, a clear coherence peak was observed just below $T_{\rm c}$ up to 2.2~GPa (phase III), indicating conventional s-wave superconductivity.
%In contrast, the coherence peak disappeared at 3.1~GPa (phase IV), indicating that the SC gap symmetry changes in phase IV and it may be related to the predominance of a spin-triplet channel caused by the increase in parity mixing.

\section*{Acknowledgments}
The authors acknowledge S. Yonezawa, Y. Maeno and M. Takigawa for fruitful discussions. 
This work was partially supported by the Kyoto University LTM Center and Grant-in-Aids for Scientific Research (KAKENHI) (Grants No. JP15H05882, No. JP15H05884, No. JP15H05745, No. JP17K14339, No. JP18H04323, No. JP19K14657, and No. JP19H04696).

%\bibliographystyle{apsrev4-1}
%\bibliographystyle{jpsj}
%\bibliography{Ref,NMR}

\end{document}